\documentclass[final,3p,sort&compress]{elsarticle-arxiv}

\usepackage{graphicx}

\usepackage{color,amsmath,amssymb,amsfonts,ulem}
\usepackage{bm}

\def\be{\begin{equation}}
\def\ee{\end{equation}}
\def\ba{\begin{eqnarray}}
\def\ea{\end{eqnarray}}
\def\bd{\begin{displaymath}}
\def\ed{\end{displaymath}}
\def\bq{\begin{eqnarray}}
\def\eq{\end{eqnarray}}

\begin{document}

\begin{frontmatter}



\title{What dynamics can be expected for mixed states in two-slit
experiments?}


\author[label1]{Alfredo Luis}
\author[label2]{\'{A}ngel S. Sanz\corref{corresp}}
\ead{asanz@iff.csic.es}

\cortext[corresp]{Corresponding author}

\address[label1]{Departamento de \'Optica, Universidad Complutense de
Madrid, 28040 Madrid, Spain}

\address[label2]{Instituto de F\'{\i}sica Fundamental (IFF-CSIC),
Serrano 123, 28006 Madrid, Spain}

\begin{abstract}
Weak-measurement-based experiments [Kocsis {\it et al.}, Science
332 (2011) 1170] have shown that, at least for pure states, the
average evolution of independent photons in Young's two-slit experiment
is in compliance with the trajectories prescribed by the Bohmian
formulation of quantum mechanics.
But, what happens if the same experiment is repeated assuming that the
wave function associated with each particle is different, i.e., in the
case of mixed (incoherent) states?
This question is investigated here by means of two alternative
numerical simulations of Young's experiment, purposely devised
to be easily implemented and tested in the laboratory.
Contrary to what could be expected a priori, it is found that even
for conditions of maximal mixedness or incoherence (total lack of
interference fringes), experimental data will render a puzzling and
challenging outcome: the average particle trajectories will still
display features analogous to those for pure states, i.e.,
independently of how mixedness arises, the associated dynamics is
influenced by both slits at the same time.
Physically this simply means that weak measurements are not able to
discriminate how mixedness arises in the experiment, since they only
provide information about the averaged system dynamics.
\end{abstract}



\end{frontmatter}



\section{Introduction}
\label{sec1}

Since the inception of quantum mechanics our understanding of quantum
systems is essentially based on two intertwined principles: uncertainty
and complementarity.
This landscape started changing in 2011, with experiments showing that
it is possible to reconstruct the photon wave function from direct
measurements \cite{lundeen:Nature:2011} and to infer how photons travel
(on average) in Young's experiment \cite{kocsis:Science:2011}.
This apparent breach in the above principles is possible through
the experimental implementation of the concept of weak measurement
\cite{aharonov:PRL:1988,aharonov:PRA:1990,dressel:arxiv:2013}, which
does not constitute a true violation of the quantum rules, but only
looking at them with different eyes.
Strong (von Neumann) measurements lead the measured system to
irreversibly collapse onto one of the pointer states of the measuring
device.
On the contrary, weak measurements only produce a slight perturbation
on the system, which may still continue as an almost unaltered
evolution.
Consequently, the pointer of the measuring device only undergoes a
slight deviation.
This information, together with the one arising from a subsequent
strong measurement, is enough to completely specify the state of the
system.
That is, the system quantum state can be determined from a single
experiment just by measuring the probability density and its
transversal flow (accounted for by the current density), unlike
other traditional methods, such as quantum state tomography
\cite{vogel,raymer,mukamel,wineland}, which require several
complementary experiments in order to obtain a full picture of the
corresponding quantum state.
Rigorously speaking, if $|\phi_i\rangle$ and $|\phi_f\rangle$ denote
pre- and post-selected states of the system, respectively, the weak
value rendered by a weak measurement associated with an operator
$\hat{A}$ is defined as
\be
 A_w \equiv \frac{\langle\phi_f|\hat{A}|\phi_i\rangle}
       {\langle\phi_f|\phi_i\rangle} .
 \label{eq0}
\ee
Given the dependence of the weak value $A_w$ on the system pre- and
post-selected states, it can be cleverly enhanced by choosing these
states in such a way that they approach the orthogonality.

One of the targets of the weak-measurement technique has been the
Bohmian formulation of quantum mechanics
\cite{leavens:FoundPhys:2005,wiseman:NewJPhys:2007,mir:NewJPhys:2007,
matzkin:PRL:2012,schleich:PRA:2013,bliokh:NewJPhys:2013,braverman:PRL:2013},
also known as Bohmian mechanics, due to the impossible but appealing
idea of dealing with well-defined trajectories in quantum mechanics.
This quantum approach is a way to recast quantum mechanics in terms of
a hydrodynamic language, equivalent to the formulations formerly
proposed by Schr\"odinger, Heisenberg, Dirac, Feynman, Wigner, etc.
In Bohmian mechanics \cite{madelung:ZPhys:1926,bohm:PR:1952-1,sanz-bk-1}
the flow of the probability density in configuration space, which
accounts for the system time-evolution, is monitored by means of
streamlines or trajectories. Within this scenario weak values are
directly connected to the Bohmian momentum as
\cite{wiseman:NewJPhys:2007,hiley:JPhysConfSer:2012}
\be
 {\bf p} = \nabla S =
  {\rm Re}\left[\frac{\langle {\bf r}|\hat{\bf p}|\psi(t)\rangle}
   {\langle {\bf r}|\psi(t)\rangle}\right]  ,
 \label{eq1}
\ee
where $S({\bf r},t)$ is the phase of the system wave function when the
latter is recast in polar form,
\be
 \psi({\bf r},t) = \rho^{1/2}({\bf r},t) e^{iS({\bf r},t)/\hbar} ,
 \label{pa}
\ee
with $\rho ({\bf r},t) = | \psi ({\bf r},t) |^2$ being the probability
density.
Actually, the momentum ${\bf p}$ can be rigorously placed within the
standard Schr\"odinger picture in coordinate representation as the
quotient between the probability density current and the probability
density (see Section~\ref{sec22} below).
The expression between square brackets on the right-hand side thus
corresponds to the weak value associated with a (weak) measurement of
the usual momentum operator, $\hat{\bf p} = -i\hbar\nabla$, followed by
a standard (strong) measurement of the position operator $\hat{\bf r}$
(with outcome ${\bf r}$).
The combination of the two measurements eventually provides the value
of the local momentum ${\bf p}$ at the position ${\bf r}$.
In particular, the quantity measured in the experiment performed by
Kocsis {\it et al.}~\cite{kocsis:Science:2011} was the transversal
component of this momentum.

Moved by the fact that Bohmian trajectories are in compliance with
the experimental data reported in \cite{kocsis:Science:2011}, which
provide an intuitive picture of the average dynamical evolution of
a swarm of individual photons in Young's experiment, it seems to be
a very natural question to ask about the quantum dynamics associated
with mixed states in an analogous situation.
In this regard, if the same experiment is repeated introducing some
degree of incoherence, the collection of frames of the transverse
momentum will give us an idea on how the corresponding mixed state
evolves, even though we are aware that such a state only represents
a kind of average information about the quantum system.
Leaving aside the question about what such a state really represents or
if it really exists, our purpose here is just to investigate which kind
of trajectories can be associated with it and, therefore, what could be
expected from a real experiment, which in principle could be easily
tested by means of the same experimental setup used in
\cite{kocsis:Science:2011}, although including some minor changes
related to the way how the mixedness or incoherence is treated.

Specifically, to that end we have considered two simple numerical
implementations of Young's experiment, which represent two possible
ways to reach the same mixed state.
In one of these scenarios, one and only one slit is randomly open at a
time, while in the other each time that the particle passes through the
two slits, an extra random phase arises between the two diffracted
waves.
In both cases, the final outcome is the same: interference fringes are
removed.
Notice that in the first scenario one would expect the classical-like
result of trajectories for two independently open slits, while in the
second scenario trajectories evolve with the information that both
slits are open.
Now, under such circumstances, how does the transverse momentum (weak
value) looks like if we perform measurements at different positions
from the slits as in \cite{kocsis:Science:2011}?
What we have noticed is that the trajectories associated with sets
of measurements render in the end results that display analogous
properties to those of the pure case, namely that trajectories avoid
crossing through the same point at the same time even if we do not
observe any interference features.
In this regard, notice that even if the density matrix only describes
lack of knowledge on some experimental conditions (which slit has been
randomly open, or which random phase has been added), thus representing
an epistemic state of the system, weak measurements cannot discriminate
how the experiment is performed, thus providing us with information on
the whole setup (both slits open at a time), just as if we had an ontic
state.

The work has been organized as follows.
In Section~\ref{sec2} we explore the guidance equation for mixed states
$\hat{\rho}$ in a Young-type interferometer, emphasizing its nonlinear
nature by considering two different physical realizations of
$\hat{\rho}$.
In Section~\ref{sec3} we propose a practical experiment revealing the
nonclassical features of the guidance equation in the case of totally
mixed $\hat{\rho}$ leading to no interference effect.
The main outcomes from this work as well as some related remarks are
summarized in Section~\ref{sec4}.


\section{Mixed state dynamics: Theoretical approach}
\label{sec2}


\subsection{State preparation and expected measured outcomes}
\label{sec21}

Let us consider the passage of a particle, such as a photon or an
electron, through the two slits of a Young-type interferometer.
For a better understanding of the approach described below, it is
illustrative briefly revisiting the basic physical ideas behind the
original experiment.
To that end, it is enough to start the description once particles (in
this case, photons) have been diffracted by the two slits; more details
on how the experiment was exactly performed can be found in the original
work \cite{kocsis:Science:2011}.
Thus, essentially, once the distance between the slits and the detector
is fixed (by adjusting a set of lenses), the latter collects the flux
of photons, one by one, for a certain exposure time and at different
positions along the direction parallel to the slits (the transversal
direction).
So far, this step is standard in experiments of this kind.
In order to determine within the same experiment the average transverse
momentum and, therefore, get information about the photon average flow
along the transversal direction, a polarizer that produces a slight
variation of the photon polarization state (weak measurement) is
introduced at a certain distance behind the two slits.
Notice in this procedure that the information obtained is statistical,
because for each position of the detector it is needed a physically
significant number of photons (directly related to the detector exposure
time).
In this sense, at each distance from the slits, the intensity pattern
is proportional to the number of photons that reach a given position
in a certain (exposure) time and, more importantly, the associated
transverse momentum corresponds to the average (transverse) momentum
of those photons.
Therefore, the trajectories or paths inferred from the experimental
data recorded only provide us with information on the average
(transverse) flow associated with a large ensemble of independent
photons, saying nothing about the individual motion of each photon,
which remains unknown.
This is the maximum amount of information that the experiment will
render, thus avoiding the violation of any fundamental quantum
principle.

Note that in the above experiment one can assume without loss of
generality that the source is highly coherent, so that the wave
function associated with the particle before reaching the slits is
nearly monochromatic, i.e., a plane wave, for practical purposes.
This is rather common working hypothesis in matter wave interferometry
as well as in its optical counterpart (which is, nonetheless, based on
experimental evidence, of course).
Now, let us consider exactly the same experimental setup, including
the same statistical procedure.
However, after the particle gets diffracted and before it reaches the
polarizer, we assume that the two outgoing probability amplitudes (one
for each slit) are acted in such a way that they no longer constitute
a coherent superposition.
This results in a mixed state described by a density matrix
$\hat{\rho}$, which does not give rise to interference features if
it is maximally mixed.
There are two physical scenarios compatible with this $\hat{\rho}$,
which are associated with two different ways to carry out the
experiment:
\begin{itemize}
\item[i)] This state may arise from a situation where one and only one
slit is randomly open at a time.
For example, each slit can be followed by a shutter; both shutters are
somehow connected to and controlled by a relay, so that when one of the
shutters is open the other is closed, and vice versa.
The open/close procedure is automatic and random, and takes place before
the photon has reached the slits.
In this way, if we do not keep track on which shutter (slit) is open at
each time, the mixed state describing the experiment is given by
\be
 \hat{\rho} (t)  =
  \sum_\lambda p_\lambda
   |\psi_\lambda (t) \rangle\langle\psi_\lambda (t) |
   = p_+ |\psi_+ (t) \rangle\langle\psi_+ (t) |
   + p_- |\psi_- (t) \rangle \langle \psi_- (t) |  ,
 \label{f1}
\ee
where $|\psi_\lambda (t) \rangle$ denotes the probability amplitude
coming from slit $\lambda$ ($\lambda=\pm$), with $p_\lambda$ being
positive real numbers such that $p_+ + p_-  =1$ (for maximal mixedness
$p_+ = p_- = 1/2$).
Formally speaking, Eq.~(\ref{f1}) appeals to a physical picture where
each system realization is prepared either in state $\psi_+$ or
$\psi_-$, with probabilities $p_+$ and $p_-$, respectively
---$\psi_+$ and $\psi_-$ never coexist and therefore we never see
interference at a distant screen [notice the lack of coherence terms
$|\psi_\pm (t) \rangle\langle\psi_\mp (t) |$ in (\ref{f1})].

\item[ii)] Alternatively, mixedness can also be produced by adding a
random, uncorrelated constant phase $\delta$ between the two diffracted
probability amplitudes.
From an experimental viewpoint, this would correspond to a situation
where one of the slits is followed, for example, by a polarizer such
that its polarization axis may acquire a random orientation before
the photon reaches the two slits.
In this case, the discrete sum of (\ref{f1}) becomes an integral in
order to account for the continuous phase value, between 0 and $2\pi$,
and consequently $\hat{\rho}$ reads as
\be
 \hat{\rho}(t) = \int_{2 \pi} d \delta  P(\delta )
  |\psi_\delta  (t) \rangle \langle \psi_\delta (t) | ,
 \label{f2}
\ee
where
\be
 |\psi_\delta (t) \rangle = \sqrt{p_+} | \psi_+ (t) \rangle
  + e^{i \delta} \sqrt{p_-} |\psi_- (t) \rangle  ,
\ee
and $P(\delta) $ is the statistics for $\delta$, with $P(\delta) =
(2\pi)^{-1}$ for maximal mixedness.
As before, because the different values of $\delta$ are unknown, at
each distance from the slit the probability density is just an average
over all the realizations.
Hence, although for each particular value of $\delta$ there is always
interference at a distant screen, with visibility $V_\delta = 2
\sqrt{p_+ p_-}$, the eventual interference fringes are washed out
due to the $\delta$-averaging.
\end{itemize}

In principle, regardless of the interpretation that can be ascribed to
$\hat{\rho}$, experiments such as the above described ones would allow
to monitor the time-evolution of the corresponding (averaged)
probability density by determining it at different distances from the
slits, just as in the experiment reported in \cite{kocsis:Science:2011}.
These experiments would not be able to discriminate between the two
scenarios, thus rendering the same average quantities, a signature that
mixedness can be reached in many physically different but formally
equivalent ways [of course, unless one keeps records of which shutter
is open at each time, but then we would have separately
$\hat{\rho}_\pm (t) = |\psi_\pm (t) \rangle\langle\psi_\pm (t)|$,
instead of (\ref{f1})].
Furthermore, it is also worth stressing that in these experiments
coherence times do not play any role in principle, unlike experiments
involving loss of coherence by decoherence, for instance.
In the above experiments it is assumed that the passage of particles
(e.g., photons, as in \cite{kocsis:Science:2011}) is performed in
such a way that there is one and only one particle crossing the
interferometer at each time.
In this regard, the suggested experiments would be close in spirit to
the one recently carried out by Matteucci {\it et al.}~\cite{matteucci}
consisting in reproducing Young's interference pattern with electrons
launched in such a way that there is no possibility that these particles
can be time-correlated.
This means that the only relevant time-scales would be the time span
between two consecutive events and the time needed to (randomly) change
the state of the shutter or the polarizer.
Now, since both experiments can be in principle performed leaving
a long delay between consecutive experiments and launching each
individual particle at will (providing we do not look at the state
of shutters/polarizers), the relevance of these time-scales is also
relative.


\subsection{Guidance equation for mixed states}
\label{sec22}

The weak value associated with the momentum operator (to be more
precise, its transverse component) corresponds to the Bohmian momentum
or guidance equation, which is well defined for pure states.
To obtain a generalized version, also valid for mixed states, let us
start from the von Neumann equation,
\be
 \frac{\partial \hat{\rho}}{\partial t}
  = -\frac{i}{\hbar} \left[ \hat{H},\hat{\rho} \right] ,
 \label{vN1}
\ee
which describes, in general, the time-evolution of the density matrix
$\hat{\rho}$.
In the position representation, the elements of the density matrix read
as $\rho({\bf r}',{\bf r},t) = \langle {\bf r}'|\hat{\rho}(t)|{\bf r} \rangle$;
the probability density $\rho({\bf r},t)$ comes from the diagonal of
$\rho({\bf r}',{\bf r},t)$ (i.e., for ${\bf r}' = {\bf r}$).
The time-evolution of the probability density associated with a mixed
state can be obtained from Eq.~(\ref{vN1}) by projecting both sides of
this equation onto the position space.
This results in
\be
 \frac{\partial \rho({\bf r},t)}{\partial t}
  = -\frac{i}{\hbar}\ \!
  \langle {\bf r}| \! \left[ \hat{H},\hat{\rho} \right]
  \! |{\bf r}\rangle .
 \label{vN2}
\ee
After some algebra, and taking into account that
$\langle {\bf r} | \nabla | {\bf r}' \rangle = \nabla_{\bf r}
\delta ({\bf r} - {\bf r}')$, we find the continuity equation
\be
 \frac{\partial \rho({\bf r},t)}{\partial t}
  = - \nabla \cdot {\bf J}({\bf r},t) ,
 \label{vN3}
\ee
where
\be
 {\bf J} = \frac{1}{m}\ \!
  {\rm Im} \left[ \nabla_{{\bf r}'} \rho ({\bf r}',{\bf r},t)
  \right]_{{\bf r'} = {\bf r}}
 \label{J1}
\ee
is the current density associated with $\rho({\bf r},t)$.
Nonetheless, in general, regardless of the chosen representation, the
current density can be expressed as \cite{Cohen}
\be
 {\bf J} ({\bf r},t) = \mathrm{tr} \left[ \hat{\rho} (t) \hat{K}
 ({\bf r}) \right] ,
 \label{pdc}
\ee
where
\be
 \hat{K} ({\bf r}) = \frac{1}{2m} \left(
 | {\bf r} \rangle \langle {\bf r} | \hat{\bf p} + \hat{\bf p}  |
 {\bf r} \rangle \langle {\bf r} | \right)
\ee
is the current density operator.
In analogy to transport phenomena satisfying a conservation law
like Eq.~(\ref{vN2}), and taking into account the functional form
(\ref{pdc}), we can assume that ${\bf J}$ represents an advective flux
associated with a velocity field
\be
 \dot{\bf r}_{\hat{\rho}} ({\bf r},t) =
 \frac{{\bf J} ({\bf r},t)}{\rho ({\bf r},t)} ,
 \label{ge0}
\ee
with the subscript $\hat{\rho}$ denoting the fact that the trajectory
is obtained with the information supplied by this density matrix (for
pure states we shall simply use $\rho$).
Physically, this velocity field describes how the probability density
$\rho({\bf r},t)$, taken as a scalar field, is transported throughout
the position space.
If only the density matrix in the position representation,
$\rho({\bf r}',{\bf r},t)$, is accessible, after Eq.~(\ref{J1}) we may
then use \cite{sanz:EPJD:2007}
\be
 \dot{\bf r}_{\hat{\rho}} ({\bf r} ,t) = \frac{\hbar}{m}\ \! {\rm Im}
  \left[ \frac{{\bf \nabla}_{{\bf r}'}
  \rho({\bf r}',{\bf r},t)}{\rho({\bf r}',{\bf r},t)}
  \bigg\arrowvert_{{\bf r}'={\bf r}} \right] .
 \label{eq3}
\ee

For pure states, ${\bf J}$ acquires the usual form
\be
 {\bf J} ({\bf r},t) = \frac{\hbar}{m}\ \! \mathrm{Im}\left [
 \psi^\ast  ({\bf r},t)  {\bf \nabla}  \psi ({\bf r},t) \right ]
 = \frac{1}{m}\ \! \rho ({\bf r},t) {\bf \nabla} S  ({\bf r},t) ,
\ee
where we have made use of the polar ansatz (\ref{pa}) in the last
equality.
Accordingly, Eq.~(\ref{eq3}) reduces to the usual Bohmian guidance
equation
\be
 \dot{\bf r}_{\rho} ({\bf r},t) =
 \frac{1}{m}\ \! {\bf \nabla} S ({\bf r},t) .
\ee
On the other hand, if the mixed state is represented as
\be
 \hat{\rho} = \sum_\lambda p_\lambda | \psi_\lambda \rangle
  \langle \psi_\lambda | ,
 \label{ed}
\ee
and the current density as
\be
 {\bf J} ({\bf r},t) =\sum_\lambda p_\lambda {\bf J}_\lambda
 ({\bf r},t) = \frac{1}{m}
   \sum_\lambda p_\lambda \rho_\lambda ({\bf r},t)
     \nabla S_\lambda  ({\bf r},t) ,
 \label{eqJ}
\ee
where each $\psi_\lambda$ itself has been recast in polar form
in the last equality, i.e., $\psi_\lambda = \sqrt{\rho_\lambda}
e^{i S_\lambda /\hbar}$, the resulting guidance condition reads as
\be
  \dot{\bf r}_{\hat{\rho}} ({\bf r},t) = \frac{\sum_\lambda p_\lambda
  \rho_\lambda ({\bf r},t) \nabla S_\lambda  ({\bf r},t)} {m
  \sum_\lambda p_\lambda \rho_\lambda ({\bf r},t) }
  = \frac{\sum_\lambda p_\lambda
    \rho_\lambda ({\bf r},t) \dot{\bf r}_\lambda}
     {\sum_\lambda p_\lambda \rho_\lambda ({\bf r},t) } ,
 \label{ge}
\ee
where
\be
 \dot{\bf r}_\lambda ({\bf r},t) =
  \frac{1}{m}\ \! \nabla S_\lambda ({\bf r},t) .
  \label{ge2p}
\ee
The guidance equation (\ref{ge}) is nonlinear in both the probabilities
$p_\lambda$ and the probability densities $\rho_\lambda ({\bf r},t)$,
thus clearly providing us with different predictions with respect to
the simple average quantity
\be
 \bar{\bf v} \equiv \dot{\bf r}_{\rm av} ({\bf r},t) =
  \sum_\lambda p_\lambda \dot{\bf r}_\lambda ({\bf r},t) ,
 \label{ge2}
\ee
which one might expect a priori to be the correct outcome.
The same discussion holds for the continuous case with $\delta$,
but substituting the sum by an integral, as pointed out in
Section~\ref{sec21}.
In any case, to some extent Eq.~(\ref{ge}) would agree with an ontic
understanding of the process, while Eq.~(\ref{ge2}) would be closer to
an epistemic viewpoint.
Now, given that the information obtained for each $\lambda$ value is
available without altering the state $\hat{\rho}$, the experimenter
could use it to reconstruct the trajectories following Eq.~(\ref{ge})
or Eq.~(\ref{ge2}).
Clearly, both equations cannot be valid.
Which equation will be then in agreement with the experimental data
and, therefore, will provide us with the correct dynamical evolution
of the mixed state?


\subsection{Expected Bohmian outcomes}
\label{sec23}

Following a standard procedure, i.e., only observing the behavior of
the density matrix, both processes (i) and (ii) are compatible with
the eventual result represented by $\hat{\rho}$ via a final averaging
procedure.
However, if we also focus on the dynamical behavior of the density
matrix, i.e., its flow in configuration space, things become
different. It is at this point where the Bohmian guidance equation
(\ref{ge}) constitutes a suitable working tool: its transversal
component is proportional to the transverse flow, which can be
experimentally detected by means of weak measurements. The
possibility to somehow measure the transversal flow allows us to
test whether the flow associated with the ensemble $\hat{\rho}$  is
classical-like or not by inspecting the reconstructed trajectories.

Consider the guidance equation (\ref{ge}) associated with $\hat{\rho}$.
For simplicity and without loss of generality, we can focus on the
transversal direction thus reducing the dimensionality of the problem
to 1.
As mentioned above, this equation is not linear with respect to the
pure sets $\{|\psi_\lambda\rangle\langle\psi_\lambda|\}_{\lambda=\pm}$
or, analogously, $\{|\psi_\delta\rangle \langle\psi_\delta|\}_{\delta \in [0,2\pi)}$,
unlike Eqs.~(\ref{f1}) and (\ref{f2}), respectively.
As also seen above, $\dot{x}_{\hat{\rho}} (x,t)$ is not the average
velocity corresponding to independent random-slit velocities
$\dot{x}_\lambda (x,t)$ \cite{sanz:EPJD:2007,sanz:JPA:2008,sanz:CPL:2009-2},
\be
 \dot{x}_{\hat{\rho}} (x,t) \neq
   p_+ \dot{x}_+ (x,t) + p_- \dot{x}_- (x,t) = \dot{x}_{\rm av} ,
 \label{eq4}
\ee
with
\be
 \dot{x}_\lambda  (x,t) =
  \frac{{\rm Im} \left[ \psi^*_\lambda (x,t)
   \partial_x \psi_\lambda (x,t) \right]}{\rho_\lambda (x,t)} ,
 \label{eq5}
\ee
and where $\psi_\lambda (x,t) = \langle x|\psi_\lambda (t)\rangle$ are
the corresponding wave functions (regarding notation, $\partial_x
\equiv \partial/\partial x$).
Similarly, in the case of phase-averaging, $\dot{x}_{\hat{\rho}} (x,t)$
is not the average of random-phase velocities $\dot{x}_\delta (x,t)$,
\be
 \dot{x}_{\hat{\rho}} (x,t) \neq
  \int_{2 \pi} d \delta P(\delta) \dot{x}_\delta (x,t)
  = \dot{x}_{\rm av} ,
 \label{eq4int}
\ee
with
\be
 \dot{x}_\delta (x,t) =
  \frac{{\rm Im} \left[\psi^*_\delta (x,t)
   \partial_x \psi_\delta (x,t)\right]}{\rho_\delta (x,t)} ,
 \label{eq7}
\ee
and where $\psi_\delta (x,t) = \langle x|\psi_\delta (t)\rangle$.
Instead, according to Section~\ref{sec22}, we have
\be
 \dot{x}_{\hat{\rho}} (x,t) =
  \frac{p_+ \rho_+ (x,t) \dot{x}_+(x,t)
   + p_- \rho_- (x,t) \dot{x}_-(x,t)}
    {p_+ \rho_+ (x,t) + p_- \rho_- (x,t)}  ,
 \label{eq6}
\ee
and
\be
 \dot{x}_{\hat{\rho}} (x,t) =
  \frac{\int_{2 \pi} d \delta P(\delta) \rho_\delta (x,t)
   \dot{x}_\delta (x,t)}
  {\int_{2 \pi} d \delta P(\delta) \rho_\delta (x,t)},
 \label{eq6d}
\ee
respectively, with $\rho_\lambda (x,t) = |\psi_\lambda (x,t)|^2$ and
$\rho_\delta (x,t) = |\psi_\delta (x,t)|^2$.

The difference between Eqs.~(\ref{eq6}) and (\ref{eq6d}), and the bare
averages, given by Eqs.~(\ref{eq4}) and (\ref{eq4int}),
respectively, consists in assuming or not, respectively, a properly
weighted value of the corresponding velocity fields ($\dot{x}_\lambda$
or $\dot{x}_\delta$), with the weights including both the slit
probabilities [$p_\lambda$ and $P(\delta)$] and the respective
(continuous) probability densities ($\rho_\lambda$ and $\rho_\delta$).
The corresponding trajectories will therefore contain information
from both slits or from all phase-shifts, which will influence
importantly the outcome that we may infer from a real experiment
on mixed states.
In other words, for Eq.~(\ref{ge2}) to be valid, the $p_\lambda$ have
to be substituted by the more appropriate weights
\be
 p'_\lambda ({\bf r},t) = \frac{p_\lambda \rho_\lambda ({\bf r},t)}
  {\sum_\lambda p_\lambda \rho_\lambda ({\bf r},t)} ,
\ee
which leads to Eq.~(\ref{ge}).
Analogously, in the continuous case, accounted for by the right-hand
side of the inequality Eq.~(\ref{eq6}) in one dimension, we would need
to substitute $P(\delta)$ by
\be
 P'(\delta)({\bf r},t) = \frac{P(\delta)\rho_\delta({\bf r},t)}
  {\int_{2\pi} d\delta P(\delta)\rho_\delta({\bf r},t)} .
\ee
This requirement is demanded by the own structure of the flows
associated with each diffracted wave, which contain information about
the particular way how particles statistically distribute when they
pass through each slit.
These facts are discussed in more detail in next section.


\section{Numerical experiments}
\label{sec3}


\subsection{Scenario 1}
\label{sec31}

\begin{figure}[t]
 \begin{center}
 \includegraphics[width=15cm]{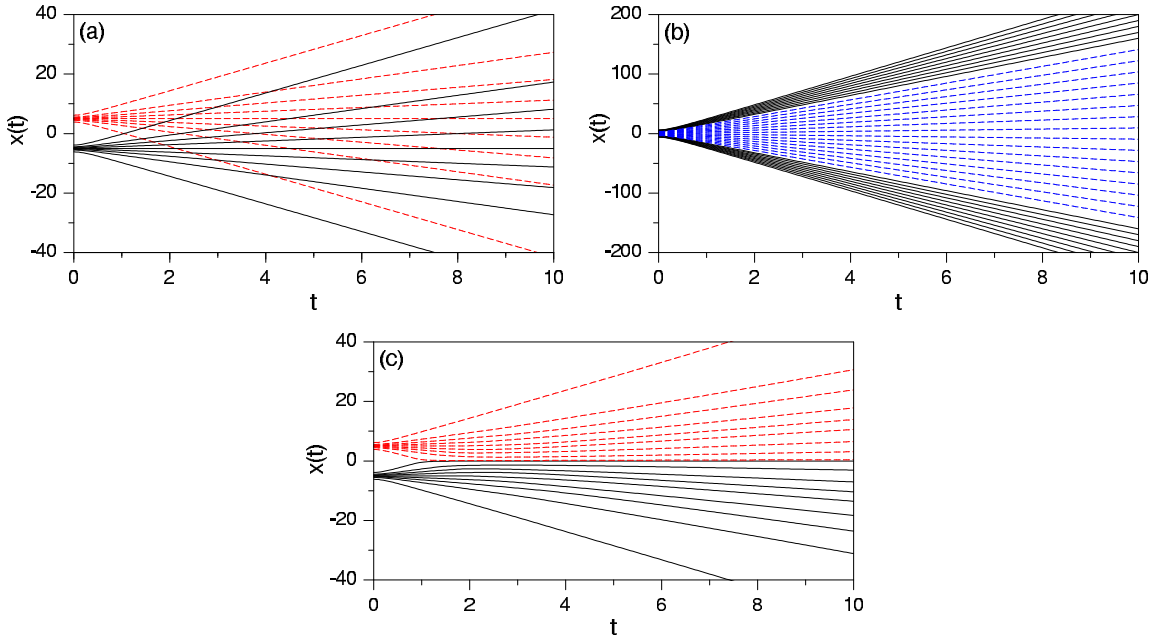}
 \caption{\label{fig1}
  Bohmian trajectories associated with: (a) slits not simultaneously
  open, (b) the bare mixture described by the right-hand side of
  Eq.~(\ref{eq4}), and (c) the weighted mixture described by
  Eq.~(\ref{eq6}).
  In panels (a) and (c), the trajectories leaving the upper slit are
  denoted with red dashed line in order to distinguish them from those
  leaving the lower one (black solid line).
  The value of the parameters used in these simulations (see text for
  details) are: $\hbar=1$, $m=0.5$, $\sigma_0=0.5$, and $d=10$.
  The initial conditions along the slits (nine for each slit) follows
  a Gaussian distribution, in agreement with $\rho_\lambda(x,0)$.}
 \end{center}
\end{figure}

We shall start our analysis from the physical realization (i), assuming
that a particle (say an electron) is sent through the interferometer
where one of the slits is blocked. According to the traditional picture,
the state of the electron is either $| \psi_+ \rangle$ or
$| \psi_- \rangle$ depending on the slit blocked.
An experiment to elucidate the suitability of the relation (\ref{eq4})
should not be much more complicated to carry out than the one reported
in \cite{kocsis:Science:2011}.
To this end the transversal momentum should be measured at a point $x$
of the observation plane with only one slit open at a time, keeping record
of which slit is open. In order to reproduce the experiment in a numerical
fashion, let us mimic the experiment carried out by Kocsis
{\it et al.}~\cite{kocsis:Science:2011,dimic:PhysScr:2013}, assuming that
\be
 \psi_\pm (x) \propto e^{-(x \pm x_0)^2/4\sigma_0^2} ,
 \label{eq8}
\ee
with the centers of the slits separated at a distance $d$ ($x_0=d/2$),
and each slit transmitting on average a Gaussian-like beam of width
$\sigma_0$ from an incoming, coherent plane wave. If the forward
propagation speed (perpendicular to the plane where the slits are positioned,
at $z=0$) is faster than the transversal diffraction, the degrees of freedom
describing these two directions can be decoupled\footnote{Some
numerical simulations have been run in order to verify this commonly
used assertion as well as the hypothesis of the Gaussianity of the
outgoing beams.}.
Then the evolution of $\psi_\pm (x)$  is given by \cite{sanz-bk-2}
\be
 \psi_\pm (x,t) \propto e^{-(x \pm x_0)^2/4 \sigma_0 \tilde{\sigma}_t} ,
 \label{evG}
\ee
where $\tilde{\sigma}_t = \sigma_0[1 + (i\hbar t/2m\sigma_0^2)]$, and
$t$ represents the propagation time (which is essentially proportional
to the distance between the plane of the slits and the observation
plane).
Since the state is always either $|\psi_+ \rangle$ or $|\psi_- \rangle$,
at each point $(x,t)$ we can naturally consider the two transverse
momenta corresponding to just one slit open, given by
\begin{equation}
\label{xpm}
\dot{x}_\pm (x,t) = \frac{\dot{\sigma}_t}{\sigma_t} \left ( x \pm x_0 \right ) ,
\end{equation}
where  $\sigma_t = |\tilde{\sigma}_t |$. The corresponding trajectories are plotted  in
Fig.~\ref{fig1}(a). When the lower slit is blocked for example, the trajectories emerging
from the upper slit can reach the $x<0$ region, and vice versa. These trajectories do
not satisfy the well-known Bohmian non-crossing property because they correspond
to two different values of  the which-slit variable \cite{sanz:JPA:2008}.

In order to address the dynamics of the whole ensemble as a random
succession of the states $|\psi_+\rangle$ and $|\psi_-\rangle$,
we remove the information about the slit crossed after the
experiment has been concluded. In a first approach we may consider
that a fraction $p_\lambda$ of the particles follow the upper/lower path so that
the ensemble implies the coexistence of the two families of trajectories in Fig. 1(a)
with the corresponding weights $p_\lambda$. This cannot hold since we know
that at each point there can only be one trajectory. Thus we may
then consider a naive average momentum at each point as $p_+ \dot{x}_+
(x,t) + p_- \dot{x}_- (x,t)$,  which is actually the right-hand side of
Eq.~(\ref{eq4}).
In particular, for  $p_+ = p_- = 1/2$ this reads as
\begin{equation}
 p_+ \dot{x}_+ (x,t) + p_- \dot{x}_- (x,t)= \frac{\dot{\sigma}_t}{\sigma_t} x .
 \label{eqbare}
\end{equation}
It turns out that this is the quantum flow of a single Gaussian wave packet centered at $x=0$.
The trajectories emerging within the slits are plotted in Fig.~\ref{fig1}(b) (note the different
scale in the $x$-axis). The average  $p_+ \dot{x}_+ (x,t) + p_- \dot{x}_- (x,t)$ also produces
well-defined trajectories between the two flows emerging from the slits in Fig.~\ref{fig1}(b),
but their backwards prolongation lead to starting points between the slits (dashed lines), and therefore
with null weight.

Alternatively, we may disregard the which-slit information from the very beginning,
processing all the experimental data without producing first Eq.~(\ref{xpm}).  At each
observation point $(x,t)$ this process should lead to a single value $\dot{x}_\rho (x,t) $ in
the left-hand side of Eq.~(\ref{eq4}). For $p_+ = p_- = 1/2$ this is
\begin{equation}
\dot{x}_\rho (x,t) = \frac{\dot{\sigma}_t}{\sigma_t} \left [ x - x_0 \tanh
\left ( \frac{x x_0}{\sigma_t^2} \right ) \right ] ,
 \label{eqcorrect}
\end{equation}
where we have used  Eqs.~(\ref{eq6}),  (\ref{xpm}), and   the probability densities \cite{sanz-bk-2}
\begin{equation}
\rho_\pm (x,t) \propto e^{-(x \pm x_0)^2/2\sigma_t^2} .
\end{equation}
The corresponding trajectories are displayed in Fig.~\ref{fig1}(c).

The main result of this work stems from the striking nonclassical nature of the ensemble trajectories
in Fig.~\ref{fig1}(c).  They are clearly different from the straight lines that one would expect for free
classical particles emerging from one or the other slit (we can recall that even the propagation of
the Wigner function for free particles relies on straight line propagation \cite{schleich:PRA:2013}).
The differences with classical physics go beyond diffraction since, for example, the trajectories
emerging from the upper slit never reach the space $x<0$, and  vice-versa. So the ensemble
trajectories emerging from different slits seems to repel each other, although the slits are never
open simultaneously.

We may find also striking the behavior of the trajectories in Fig.~\ref{fig1}(c) when compared with
those either in Fig.~\ref{fig1}(a) or Fig.~\ref{fig1}(b). In a naive approach, one would consider that
electrons passing through a definite slit should follow the trajectories in Fig.~\ref{fig1}(a), and the
experiment would confirm this when properly sorting the data according to the which-slit knowledge.
However,  when the same electrons are regarded as part  of an ensemble the topology of the
trajectories will be completely different. The differences between Figs. \ref{fig1}(a) and \ref{fig1}(c)
cannot be explained by classical-like non contextual averaging, as clearly shown in Fig.~\ref{fig1}(b).
The failure of the classical-like reasoning is rather surprising since we are dealing with particles
that certainly cross one slit or the other, there is no quantum interference at any stage so there
should be no quantum ``mystery'' at all \cite{sanz:JPA:2008,herrmann:AJP:2002,sanz:AnnPhysPhoton:2010}.

We would like to note that this striking behavior may also be revealed by using standard pictures
of quantum mechanics other than the Bohmian approach, even though with the hydrodynamic
language of Bohmian mechanics it manifests in a more straightforward fashion. Within more
standard pictures the same results would be obtained just in terms of the quotient of the probability
density current to the probability density. The result that we have found essentially means that the
average of such a quotient [which would lead to the right-hand side of Eq.~(\ref{eq4})] is not equal
to the weighted quotient [which would lead to the left-hand side of Eq.~(\ref{eq4})].


\subsection{Scenario 2}
\label{sec32}

\begin{figure}[t]
 \begin{center}
 \includegraphics[width=15cm]{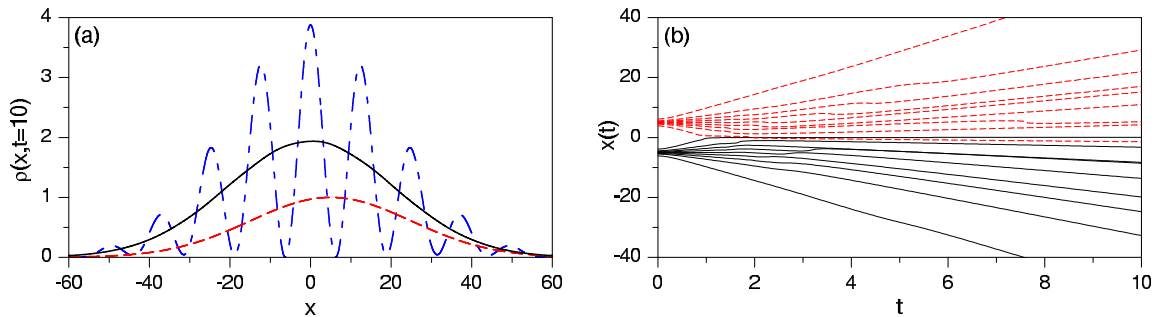}
 \caption{\label{fig2}
  (a) Probability density (black solid line) for a mixed state
  displaying total incoherence.
  This curve has been obtained after averaging over 2000 realizations
  with $\delta$ randomly chosen for each one of them.
  To compare with, the probability densities for total coherence
  ($\delta=0$; blue dash-dotted line) and for only the upper slit
  being open (red dashed line) are also included.
  (b) Sample of Bohmian trajectories associated with wave functions
  with different, randomly-chosen values of $\delta$.
  Colors (types of line) indicate trajectories leaving different slits;
  notice that because trajectories belong to different wave functions,
  they can pass through the same point at the same time (the Bohmian
  non-crossing rule does not hold in this case).
  The value of the parameters used in these simulations as well as
  the distribution of initial conditions have been taken as in
  Fig.~\ref{fig1}.}
 \end{center}
\end{figure}

Analogous features to those described in the above scenario are also
observed in the physical realization (ii), although experimentally it
is a bit more sophisticated.
To illustrate this fact, we performed a numerical simulation of Young's
experiment where for each realization (i.e., each time an electron is
sent) both slits are simultaneously open and coherent.
Now we assume that the state at the plane of the slits is
\be
 \psi(x) \propto e^{-(x-x_0)^2/4\sigma_0^2}
  + e^{ i\delta} e^{-(x+x_0)^2/4\sigma_0^2 } .
 \label{eq9}
\ee
This state evolves in time as \cite{sanz-bk-2}
\be
 \psi(x,t) \propto e^{-(x-x_0)^2/4 \sigma_0 \tilde{\sigma}_t}
  +  e^{ i\delta} e^{-(x+x_0)^2/4 \sigma_0 \tilde{\sigma}_t } ,
 \label{eq10}
\ee
where $\delta$ is some relative phase between both slits that will be
different for each realization of the experiment (i.e., each electron).
The corresponding density matrix,
$\rho(x,x',t)$, is also analytical and its diagonal corresponds to the transversal
intensity distribution,
\be
 \rho(x,t) \propto
  e^{-(x-x_0)^2/2\sigma_t^2} + e^{-(x+x_0)^2/2\sigma_t^2}
  + 2e^{-(x^2+x_0^2)/2\sigma_t^2}
  \cos \left( \frac{\hbar t}{2m\sigma_0^2}\frac{x_0 x}{\sigma_t^2}
  + \delta \right) .
 \label{eq11}
\ee
This is represented in Fig.~\ref{fig2}(a)
for $\delta=0$ (blue dashed line).
A set of Bohmian trajectories, each one corresponding to a wave
function with a different relative phase, $\delta$, is plotted in
Fig.~\ref{fig2}(b).
Because of the incoherence among wave functions, the trajectories can
cross one another, although they have nothing to do with those displayed
in Fig.~\ref{fig1}(a), where the crossings arise as a consequence of
having one slit open at a time (i.e., the
trajectories are associated with one wave packet or the other, but not
with both at the same time).

\begin{figure}[t]
 \begin{center}
 \includegraphics[width=15cm]{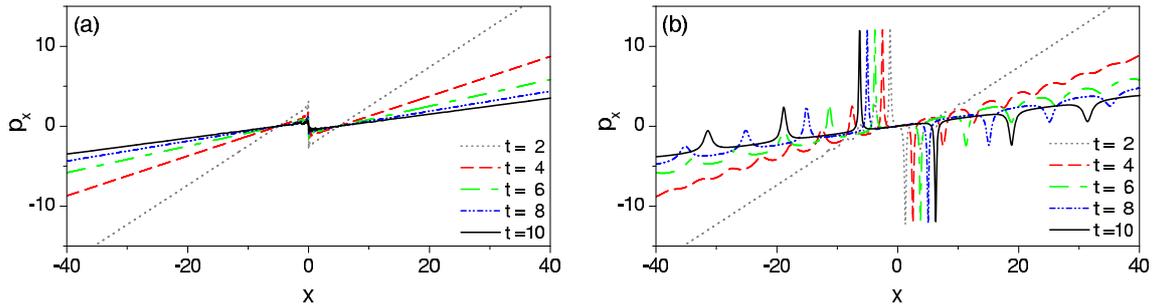}
 \caption{\label{fig3}
  (a) Space variation of the transversal momentum $p_x$ for total
  incoherence.
  Different colors (types of line) indicate different times or,
  equivalently, distances from the two slits.
  To compare with, the same function is displayed in panel (b) for
  to total coherence ($\delta=0$), evaluated at the same times
  considered in panel (a).
  The values of the parameters used in these simulations as well as
  the distribution of initial conditions have been taken as in
  Fig.~\ref{fig1}.}
 \end{center}
\end{figure}

Next we assume that any value of the phase has the same probability of occurrence,
i.e., the phase shift follows a random uniform distribution in the interval $[0,2\pi$).
As seen in Fig.~\ref{fig2}(a), when averaging over $\delta$, we find that
the intensity distribution (black solid line) is just the bare sum
of the intensities coming from each slit, with any interference feature being completely
washed out.
The intensity distribution displays a classical-like pattern (the
outcome from a typical strong measurement), but the dynamics, monitored
in terms of Bohmian trajectories (reconstructed from data collected
through weak measurements), differs from what one would expect, namely
a crossing of trajectories \cite{sanz:JRLR:2010}.
The corresponding trajectories look like those displayed in
Fig.~\ref{fig1}(c): neither they display a wiggling topology, nor they
cross one another.
The predicted transversal momentum as a function of the transversal $x$
coordinate is displayed in Fig.~\ref{fig3}(a) for different distances
(times) from the two slits.
Any trace of interference [the spiky-like behavior at periodic
values of $x$ for a given time \cite{sanz-AOP15}, as seen in
Fig.~\ref{fig3}(b)] is washed out due to the strong incoherence
leading to mixedness.
This is the trend that one should find in an experiment performed
following this route.
The experiment itself could be implemented by introducing
a small modification of the one used by Kocsis {\it et
al.}~\cite{kocsis:Science:2011}, consisting of inserting a polarizer
behind one of the slits and then randomly changing its polarization
axis at each realization.

\section{Final remarks}
\label{sec4}

Summarizing, we have gone a step beyond the experiment reported
in \cite{kocsis:Science:2011}, showing how different the dynamics of a
fully mixed state can be with respect to the preconceived picture that
we usually associate with these states.
Incoherent superpositions are commonly regarded as describing a
classical-like situation, which relies on neglecting any further inquiry
about the time evolution of the corresponding mixed state
\cite{peres:SHPMP:2002}.
In the context of our work this translates into the fact that a particle
goes through one or another slit in Young's experiment, thus avoiding any
possibility to exhibit interference features
\cite{sanz:AnnPhysPhoton:2010}.
This rather interesting and widespread conception turns out to be actually
a challenging misconception if we revisit the idea of mixed state from a
dynamical viewpoint, in particular within the Bohmian representation of
quantum mechanics, just as it was utilized in \cite{kocsis:Science:2011}
to reconstruct the average paths describing the time-evolution of single
photons in Young's experiment.
What we have shown here is that, regardless of how mixedness is
introduced, the outcome that may be expected from analogous experiments
is an average flow still governed by the presence of the two slits,
even under conditions of total lack of interference features.
This general conclusion is not only in compliance with standard quantum
pictures, but it is also in favor of alternative approaches.
For example, Gr\"ossing {\it et al.}~\cite{GFMS} have also been able to
reproduce the behavior displayed in Fig.~\ref{fig1}(c) by means of a
so-called ``superclassical'' approach, aimed at describing and explaining
quantum behaviors as an emergent phenomenon arising from the interplay
between classical boundary conditions and a classical subquantum domain
\cite{GFMS2}.

To conclude, we would like to highlight that experiments such as those
reported in \cite{lundeen:Nature:2011} or \cite{kocsis:Science:2011}
precisely show that we are still far from a full exploitation of the
possibilities offered by quantum mechanics.
Actually, experiments targeting general quantum states have recently
been proposed \cite{bamber:PRL:2012} and performed \cite{bamber:arxiv:2013}.
Our proposal here goes in this direction. If we can experimentally measure the
transversal flow, will the outcomes be in compliance with what one would expect
from two independent slits (typical classical behavior), or, on the contrary,
they will retain their quantum essence and will show an influence
from both slits at the same time?


\section*{Acknowledgements}

Support from the Ministerio de Econom{\'\i}a y Competitividad (Spain)
under Project No.~FIS2011-29596-C02-01 (AS) and FIS2012-35583 (AL),
and a ``Ram\'on y Cajal'' Research Fellowship with Ref.~RYC-2010-05768
(AS) is acknowledged.




\end{document}